\documentclass{article}
\usepackage[english]{babel}
\usepackage[utf8]{inputenc}
\usepackage{johd}
\usepackage{hyperref}
\hypersetup{colorlinks,allcolors=black}

\title{Trust and Time Preference: Measuring a Causal Effect in a Random-Assignment Experiment}

\author{Linas Nasvytis\\
        \small{Columbia University, University of Oxford} \\
}

\date{} %leave blank

\begin{document}

\maketitle

\begin{abstract} 
\noindent Large amounts of evidence suggest that trust levels in a country are an important determinant of its macroeconomic growth. In this paper, we investigate one channel through which trust might support economic performance: through the levels of patience, also known as time preference in the economics literature. Following \citet{gabaix2017myopia}, we first argue that time preference can be modelled as optimal Bayesian inference based on noisy signals about the future, so that it is affected by the perceived certainty of future outcomes. Drawing on neuroscience literature, we argue that the mechanism linking trust and patience could be facilitated by the neurotransmitter oxytocin. On the one hand, it is a neural correlate of trusting behavior. On the other, it has an impact on the brain's encoding of prediction error, and could therefore increase the perceived certainty of a neural representation of a future event. The relationship between trust and time preference is tested experimentally using the Trust Game. While the paper does not find a significant effect of trust on time preference or the levels of certainty, it proposes an experimental design that can successfully manipulate people's short-term levels of trust for experimental purposes. \end{abstract}

\section{Introduction}
Trust is a fundamental, yet also an elusive concept in economic research. As noted by Kenneth Arrow, "virtually every economic transaction has an element of trust" \citep{arrow1973information}, which helps explain why the last 30 years have witnessed a growing interest of research that examines the various ways through which trust affects economic outcomes.

On the one hand, macroeconomic studies have linked trust to economic performance of countries, starting with \citet{knack1997does}. While trust is a multidimensional variable, many studies have focused on the impact of generalized trust -- the levels of trust that people have in anonymous individuals within a society \citep{ho2021trust}. On the other hand, the introduction of the Trust Game in experimental economics by \citet{berg1995trust} has led to an increasing interest of studying the impact of trust in individual decision-making. Part of the appeal of the Trust Game lies in its simplicity of measuring trust, which allows researchers to apply the game across different cultures \citep{alos2019trust}. The game is played by two agents, Participant A and Participant B. At the beginning, Participant A is endowed with some amount of money. She can choose to send any fraction of that amount to Participant B. Before B receives the amount, it is tripled. After receiving the tripled amount, Participant B decides how much (if any) she wants to return back to Participant $\mathrm{A}$. While social preferences might play a role as a confound \citep{fehr1999theory}, the amount sent by Participant $\mathrm{A}$ is widely used as a measure of trust in an anonymous individual, while the amount of money returned measures the level of reciprocity \citep{berg1995trust}.

This paper seeks to combine both approaches to studying trust. It is primarily motivated by recent macroeconomic evidence that trust might be linked to levels of patience across countries \citep{falk2018global}. We conjecture that patience might be one of the channels through which trust affects economic growth. In particular, we hypothesize that higher levels of generalized trust might reduce the level of present bias. To understand why that might be the case, it is important to look at time preference as a consequence of noisy mental encoding of future outcomes. Specifically, we follow time discounting model from \citet{gabaix2017myopia}. The authors argue that people discount the future because it contains more noise and uncertainty, compared to the present. We believe that trust might play a role in time preference, because trust is fundamentally a belief about future outcomes. Being more trusting means being more certain about the trustworthiness about others, and thus more certain about the actions and outcomes in the future. We present evidence from biology and neuroscience on the hormone oxytocin, which supports the idea that trust might be linked to the perceived certainty of future outcomes.

To explore the potential relationship between trust and present bias, we conduct a random-assignment experiment, in which we manipulate subjects' short-term levels of trust, and then measure their levels of time preference. Levels of short-term trust are manipulated using the Trust Game. All subjects are randomly assigned to one of two treatment groups: High Trust or Low Trust. In each round, subjects are playing against a computer, who they think is another participant. In the High Trust group, the algorithm uses strategies of highly trusting and trustworthy individuals. In the Low Trust group, subjects are playing against an algorithm who is using strategies of highly untrusting and untrustworthy individuals. All strategies of the algorithm are derived from real player data in the Trust Game from \citet{fiedler2017effect}. Importantly, deception was only used because the experiment could not be conducted in a classroom setting due to COVID-19 crisis. Under normal conditions that allow for live gameplay between subjects, treatment could be implemented without any need of deception.

The experiment tests three main hypotheses. First, subjects in the High Trust group will have significantly higher levels of trust, compared to participants in the Low Trust group. Second, participants of High Trust group will have a lower level of present bias. Third, High Trust subjects will exhibit greater certainty about future outcomes.

Our results indicate that trust has no significant effect on time preference or perceived certainty about future outcomes, but that it is possible to manipulate people's short-term levels of trust for experimental purposes. The latter finding opens ample opportunities for researchers to study the causal effects of trust under experimental conditions.

The next section of this paper will provide an overview of relevant literature. The third and fourth sections provide details on experimental design and econometric specifications. The fifth section presents the main findings. The sixth and final section discusses potential problems of internal validity in the experiment, and the implications of our results.

\section{Literature Review}
\subsection{A. Trust and Time Preference: Why it matters}
To understand why studying the relationship between trust and time preference might be important, we need to start by reviewing literature on the relationship between trust and economic growth. Growing empirical evidence suggests that higher levels of trust are linked to greater macroeconomic performance of countries. \citet{knack1997does} provide evidence for a positive relationship between trust and economic growth for a sample of 29 market economies, while \citet{zak2001trust} expand the sample to 41 countries, and find the same result. More recently, using data from Global Preference Survey (GPS) across 76 countries, \citet{falk2018global} provide evidence for the same positive link between trust and economic growth.

However, the causal connection between trust and economic growth could flow either way, both ways simultaneously, or even be due to a third confound. That is why researchers have tried to establish a causal effect of trust on economic performance. \citet{knack1997does} establish this positive causal relationship by using two instrumental variables for trust - the highest "ethno linguistic" group in a society, and the fraction of law students as a percentage of all postsecondary students. However, the authors are open to admit that other confounds might be present in both of these instruments.

More recently, \citet{algan2010inherited} investigate the causal effect of trust on economic growth by focusing on the levels of inherited trust of US immigrants. For instance, by comparing trust levels of Americans with German or Italian origin, whose forebears migrated to the US between 1950-1980, the authors can detect differences in trust of these two origin countries between 1950-1980. Once they obtain the levels of inherited trust at different points in time, they can estimate the effect of a change in inherited trust levels on the change in income per capita of the countries of origin, while controlling for potential confounds. The authors find a significantly positive causal effect of trust on economic growth. Lastly, \citet{bartling2018causal} examine the causal effects of trust in an experimental setting. The authors adopt a principal-agent game with multiple equilibria, and find that trust indeed has a causal effect on the equilibrium levels of efficiency in the game.

While the current literature establishes evidence for a causal effect of trust on economic performance, it cannot easily identify the precise channels through which trust could affect economic growth. \citet{knack1997does} conjecture that higher trust could increase the efficiency of contracts, reduce reliance on formal credit institutions, and improve the quality of economic policies conducted by governmental institutions. \citet{zak2001trust} build a general equilibrium model that shows how higher trust could lead to higher increasing investments due to greater reliability of social, economic and institutional environments. However, very few empirical studies have tried to establish the validity of any of these conjectures. Even among experiment studies, only \citet{bartling2018causal} find that institutional environment appears to be the key for whether trust has causal effects and the persistence of these effects throughout time. In simple terms, our current understanding of the causal effect of trust on economic growth is similar to the way that most people understand Physics - we know it works, but we are not entirely sure how.

The first contribution of this paper is the empirical investigation of one possible channel through which trust affects economic growth -- patience. Several recent studies have found evidence for a positive relationship between trust and patience. In a sample of 76 countries, \citet{chen2013effect} measures patience as the propensity to save, and finds that individuals who think others are generally trustworthy are on average $23 \%$ more likely to have saved that year. An even more interesting finding comes from the aforementioned study by \citet{falk2018global}. The authors observe that the relationship between patience and income per capita is much stronger than the link between trust and income per capita -- in both magnitude and statistical significance. They find that patience can explain around $40 \%$ variation in the levels of income between countries. Crucially, the authors report that once trust and patience are included in a joint regression on income levels, trust loses significance. While the authors do not discuss this finding in detail, it is reasonable to conclude that trust and patience are working in similar channels to affect income levels. More specifically, trust might be working through patience to affect economic growth, which is precisely the idea that this paper will investigate.

A preliminary support for the causal effect of trust on patience comes from a study by \citet{jachimowicz2017community}, which seems to be the only paper on this topic. The authors conducted an online $2 \times 2$ experiment to establish a causal link between community trust and time discounting among low income individuals. The design involved manipulating levels of felt income (low/high), and levels of felt community trust (low/high). The results suggest that low-income individuals with higher community trust discount the future less heavily than individuals with lower community trust. While the results are encouraging, it is important to note that the study focuses on trust in local community, rather than generalized trust in anonymous individuals. In addition, the authors only focus on low income individuals, which presents a shortcoming for the extrapolation of their result to the whole population.

\subsection{Certainty: The Potential Link Between Trust and Time Preference}
The second question we need to ask is, is there a reason to believe that generalized trust might have a causal effect on time preference? To understand why that might be the case, we will turn to economic research, which investigates the reasons why time preference exists in the first place. The lion's share of literature on time preference, ranging from models of exponential discounting introduced by \citet{samuelson1937note} to quasi-hyperbolic discounting proposed by \citet{laibson1997golden}, try to answer the question of how do we discount the future, but not necessarily why. The very fact that we talk about impatience as an economic preference provides a hint that we can take it as a given, and analyze the implications for decision making assuming this preference. What all these models have in common is the assumption that people make decisions on a correct representation of rewards they receive at different dates, with complete and coherent preferences \footnote{This particular sentence is inspired by lecture notes of Michael Woodford’s “Cognitive Mechanisms and Economic Behavior” class taught in Fall 2019 at Columbia University}

More recently, researchers have been exploring the idea that time preference might not be a preference after all, but rather a cognitive illusion. Specifically, present bias might be the result of noisy mental encoding of information about future rewards. In other words, time preference could be a function of how clear we perceive the future to be compared to the present. \citet{gabaix2017myopia} propose a model of a perfectly patient Bayesian decision-maker, who receives noisy, unbiased signals about future events. Assuming that the noisiness of the future increases with distance from the present, the agent will act as if she has time preferences, even though she is simply optimizing based on posterior Bayesian beliefs about the future. In this model, discount factor is simply a function of a signal-to-noise ratio associated with the future outcome: the larger the distance between the present and the future reward, the noisier do we perceive the future reward to be. Noisiness increases the uncertainty about the future outcome, and therefore the level of present bias. Using time preference experiments, \citet{khaw2017risk} present empirical support for the idea that outcomes further in the future are indeed encoded with greater mental imprecision.

Evidence from neuroscience suggests that the key link between trust and certainty could be a neuropeptide hormone oxytocin (OT). The literature on the relation between oxytocin and trust has evolved over the last 15 years. An experimental study using the Trust Game by \citet{kosfeld2005oxytocin} found evidence for a causal link between oxytocin and trust, which has later been disputed due to problems of replication (see a review by \citet{alos2019trust}). What remains relatively convincing is the evidence that being exposed to trustworthy behavior of others in the Trust game is indeed linked to higher levels of oxytocin \citep{zak2005oxytocin}

The reason why oxytocin matters is because evidence suggests it could be linked to our certainty of the future. \citet{owen2013oxytocin} find that oxytocin enhances cortical information transfer while simultaneously lowering background activity, thus improving the clarity of signal in the brain and reducing the background noise of neurons. In the words of the authors, oxytocin increases the "signal-to-noise ratio" of information transfer, a phrase which we have encountered in the noisy mental discounting model of \citet{gabaix2017myopia}. Moreover, a review of studies on oxytocin by \citet{eskander2020neural} concludes that "what is now more widely accepted is that oxytocin has an impact on the brain's encoding of prediction error and therefore its ability to modify preexisting beliefs". If oxytocin indeed decreases the prediction error of outcomes, then it reasonable to consider the idea that oxytocin could increase our certainty about future outcomes, and therefore, the level of our time preference.

The idea that trust could be linked to certainty about the future is visible in the very definition of trust as the belief in the trustworthiness of others. As argued by \cite{ho2021trust}, it is likely that people make the decision on whether to trust someone, based on their prediction about the future trustworthiness of that person. At its very core, trust concerns the certainty about the behavior of others in the future. The second contribution of this experiment is the examination of a potential link between trust and certainty about future outcomes, in relation to time preference. To the best of our knowledge, no research has yet studied the potential link between trust and certainty.

\subsection{Experimental Methods on the Causal Effects of Trust}
The third contribution of this paper concerns experimental methods. So far, very few studies have tried to investigate the causal effect of trust in an experimental setting. We conjecture that one of the underlying reasons for that might be a lack of a reliable experimental method to manipulate people's short-term trust beliefs. \citet{bartling2018causal} have adopted a principal-agent game with multiple equilibria to study the conditions under which trust has causal effects on the equilibrium levels of efficiency. However, the experimental game seems to be specifically designed to test these findings, which makes it challenging to apply the same game to analyze causal effects of trust on other outcomes more generally.

\citet{jachimowicz2017community} conduct an online $2 \times 2$ experiment to establish a causal link between community trust and time discounting among low income individuals. The authors manipulate levels of community trust by increasing the salience of this construct in the minds of respondents. More specifically, they ask participants to either list 2 (low salience) or 10 (high salience) examples where community trust was justified. As we have mentioned before, the shortcoming of this method is twofold: first, the authors manipulate levels of community trust, which is less widely measured than generalized trust; second, the method of manipulating people's levels of trust through a questionnaire might lead to possible confounds - for example, people could potentially run out of 10 examples to justify trusting anonymous individuals.

This paper contributes to experimental literature by introducing a successful method to manipulate people's levels of generalized trust using the Trust Game. While in this particular experiment the method involves deceiving the subjects into thinking they are playing against a real participant rather than an algorithm, it is crucial to emphasize that there is no inherent need for deception, if such an experiment were conducted in a lab setting. In our case, deception was used completely out of the need to conduct the experiment online due to COVID-19 health crisis. This trust-manipulation procedure could be used in any experiment that examines the causal effect of trust on economic outcomes, or even in fields like psychology or sociology.

\section{Experimental Design}
\subsection{Overview}
The effect of trust on time preference was examined using an online experiment. While the initial design involved conducting the experiment in a classroom at Columbia University, due to COVID-19 health crisis the experiment was conducted online, using a survey platform Qualtrics.

The main procedural steps were as follows. First, subjects were provided general instructions about the experiment. Participants were told that all their answers will be anonymous. They were informed they would be playing a game with other participants in the experiment, which would be followed by a series of questions. They were also instructed about the chance to win a monetary reward, ranging from $\$ 25-40$. It was noted that the probability of winning the reward depends on their cumulative payoffs in the game. Second, trust-manipulation procedure was administered using the Trust Game. Third, subjects answered 12 time preference questions. Fourth, subjects answered 5 questions about their levels of generalized trust. Fifth, subjects answered 4 questions concerning their levels of certainty about future outcomes. Sixth, subjects responded to a series of demographic questions. Finally, subjects were provided a debriefing form about the experiment. The total duration of this procedure is around $10-15$ minutes.

The experiment was conducted in seven different online sessions, which took place on the same day, April 21, 2020. Each session was separated by two-hour intervals, the first one starting at 10:00 am, the last one starting at 10:00 pm. Every subject could only register and participate in one of the sessions. Such a design allowed to maximize the number of participants by accommodating to different time zones, while preserving the idea of a live gameplay at every session.

\subsection{Participants}
102 participants completed the experiment. They were recruited using social media. Two days before the experiment, we shared a registration form on Facebook for everyone who would like to participate in the experiment. The form described the date and time of the experiment, and asked every participant to choose one of seven time slots that they would like to participate in. The choices were 10:00-10:15 am, 12:00-12:15 pm, 4:00-4:15 pm, 6:00-6:15 pm, 8:00-8:15 pm, 10:00-10:15 pm, all in EST time. A day before the experiment, every subject was sent a reminder email. They were notified that they would be receiving the link to the experiment one minute before the beginning of their time slot. For example, if the subject had registered for 10:00-10:15 am session, they would be receiving the link at 9:59 am. The subjects were instructed to start the experiment right after receiving the link

\subsection{Trust belief-manipulation procedure}
Participants' short-term trust levels of trust were manipulated using the Trust Game. The setting of the game is as follows. When subjects were playing as Participant A, they would be endowed with $\$ 10$, and could choose to send Participant B any integer amount $x$ $(0 \leq x \leq \$ 10)$, with the hope that $\mathrm{B}$ will return some amount $y$ $(0 \leq y \leq \$ 3 x)$. When playing as Participant $\mathrm{B}$, subjects would receive some integer amount $x$, and needed to choose how much (if any) they would like to return to Participant A.

102 participants completed the experiment. They were recruited using social media. Two days before the experiment, we shared a registration form on Facebook for everyone who would like to participate in the experiment. The form described the date and time of the experiment, and asked every participant to choose one of seven time slots that they would like to participate in. The choices were 10:00-10:15 am, 12:00-12:15 pm, 4:00-4:15 pm, 6:00-6:15 pm, 8:00-8:15 pm, 10:00-10:15 pm, all in EST time. A day before the experiment, every subject was sent a reminder email. They were notified that they would be receiving the link to the experiment one minute before the beginning of their time slot. For example, if the subject had registered for 10:00-10:15 am session, they would be receiving the link at 9:59 am. The subjects were instructed to start the experiment right after receiving the link

Each participant played the game for 11 rounds. In each of these rounds, they were playing against a computer, who they thought was another participant. In the High Trust group, the algorithm used strategies of highly trusting and trustworthy individuals, derived from trust game data in \citet{fiedler2017effect}. In the Low Trust group, the algorithm was playing strategies of highly untrusting and untrustworthy individuals from the same dataset. In odd rounds (6 in total), subjects played as Participant A, while in even rounds (5 in total), they played as Participant B. Before the game began, subjects also played one practice round as Participants $B$. In this round, the algorithm sent $\$ 5$ (tripled to $\$ 15$ ) for both treatment groups. This amount is the median sent by Participant $\mathrm{A}$ in experimental data from \citet{fiedler2017effect}. It was used to avoid priming effects.

\subsection{The strategy of the computer as Participant $A$}
For the role of Participant A, High Trust strategy of the algorithm involved sending \$10, \$9, \$8 or \$7, where $\$ 10$ is the maximum amount that can be sent in the game. These amounts were sent at their respective relative frequencies observed in the data from \citet{fiedler2017effect}. In Low Trust strategy, the amount sent was either $\$ 3$, \$2, \$1 or $\$ 0$, again following their relative frequencies from the data. We chose these values for two reasons. The median amount sent in the data from \citet{fiedler2017effect} was $\$ 5$. We decided to exclude the median value and the two values right next to it, in order to create a greater contrast between the two treatments. Four different values in each treatment should also preserve the diversity of algorithm's strategies, which is needed to simulate live gameplay. A detailed account of these strategies can be found in Appendix Table 1.

\subsection{The strategy of the computer as Participant $B$}
The guiding principle for the algorithm of computer playing as Participant B is very simple: Whatever amount a real subject sends to the computer, the expected return will be (weakly) larger if she is in the High Trust treatment, and (weakly) smaller if she is in the Low Trust treatment. More specifically, for every integer amount $\mathrm{x}$ sent by a real Participant A $(0 \leq x \leq \$ 10)$, one of three amounts would be randomly returned. These three amounts depend on the treatment. Suppose the subject is playing as Participant A and sends $\$ 5$. In the High-Trust treatment, she would be returned either \$7, \$8 or \$10 with equal probability, whereas in the Low-Trust treatment, she would be returned either $\$ 2, \$ 1$ or $\$ 0$ with equal probability. The return amounts are not random - they are composed of top $25 \%$ most (least) reciprocal gameplays in the sample of \citet{fiedler2017effect}. A detailed account of these strategies can be found in Appendix Table 2.

\subsection{Note on deception}
We should note that the treatment procedure involved deceiving the subjects into thinking they are playing against a real participant, when in fact they were playing against a computer. That it is because in order to change the short-term beliefs of subjects about trusting anonymous individuals, it is necessary that they think they are interacting with one. \citet{mccabe2001cheating} have shown that in the areas of the brain usually associated with trusting behaviour -- medial prefrontal cortex (mPFC) and the temporoparietal junction (TPJ) -- increased activity during the Trust Game only takes place when subjects think they are playing against another person, rather than a computer.

The treatment procedure was carefully designed to abide by the rules that justify deception in economic experiments, set out in \citet{cooper2014note}. Cooper notes that, among other things, deception is justified when (1) the study would be prohibitively difficult to conduct without deception, and (2) subjects are adequately debriefed after the fact about the presence of deception. With regards to the first point, deception in this experiment was used only because of the need to conduct the experiment online due to COVID-19 health emergency. If the experiment was conducted in a lab setting, it would be perfectly possible to use a similar treatment without any deception. For example, all subjects could play the trust game for 3 rounds, after which they would be divided into three groups: top $25 \%$ most trusting, top $25 \%$ least trusting, and middle $50 \%$. The middle $50 \%$ participants would then be randomly assigned to play with either the very trusting or the very untrusting subjects for $\mathrm{x}$ number of rounds. This treatment should produce a very similar outcome to the treatment used in this experiment. Crucially, such a method requires live gameplay, which is extremely difficult to achieve, if subjects are conducting an online experiment at different times, as was the case in this experiment. With regards to the second point, at the end of the experiment, all subjects were also provided a debrief about how and why deception was used

\subsection{Time Preference Questionnaire}
After completing 11 rounds of trust game, all subjects were provided with a time preference questionnaire. Every subject answered 12 questions of the following form:\\

\textit{Would you rather be paid:
\begin{enumerate}
    \item \$p today
    \item \$m in $t$ days
\end{enumerate}}

Three values of $m$ \{\$25, \$30, \$40\}, and four values of t $\{1$ week, 2 weeks, 4 weeks, 8 weeks $\}$ were used. Values of $p$ were calculated based on discount rates observed in previous experiments with identical time periods in \citet{ifcher2011happiness}. All 12 pairs of binary choices a presented in Table 1.

\begin{table}
\begin{tabular}{|c|c|c|c|c|}
\hline
\multicolumn{4}{|c|}{$p$ (present value of future payment) as a function of future value $m$ and time distance $t$} &  \\
\hline
 & \multicolumn{4}{|c|}{$t$ (time period from today)} \\
\hline
$\mathbf{m}$ (future value) & $\mathbf{1}$ week & $\mathbf{2}$ weeks & 4 weeks & $\mathbf{8}$ weeks \\
\hline
$\mathbf{\$ 2 5}$ & $\$ 21$ & $\$ 21$ & $\$ 19$ & $\$ 19$ \\
\hline
$\$ \mathbf{3 0}$ & $\$ 26$ & $\$ 26$ & $\$ 23$ & $\$ 23$ \\
\hline
$\mathbf{\$ 4 0}$ & $\$ 34$ & $\$ 34$ & $\$ 31$ & $\$ 31$ \\
\hline
\end{tabular}
\caption{Present $(p)$ and future values $(m)$ of payments for different time distances $(t)$. Note; Data on discount rates, from which the present value of future payment was derived, was adopted from \citet{ifcher2011happiness}}
\end{table}

It is important to note that for every value of the future payment $m$, the present value $p$ is identical for $t \in\{1$ week, 2 weeks $\}$, and for $t \in\{4$ weeks, 8 weeks $\}$. This was used to determine the pair of $p$ and $m$ that makes a subject indifferent for each of the two time frames. The point of indifference was used to determine the discount rate for each subject. To avoid order effects, the sequence of all 12 questions, and the order of answers for each question, were randomized for each participant. In each of 12 questions, binary choices were presented in a way that maximizes the variance of responses of subjects who have different degrees of time preference. In other words, if all subjects were presented with two choices that overestimate their time preference, say $\$ 40$ today or $\$ 20$ tomorrow, we would not find much difference between the two treatment groups. A similar situation would take place, if subjects were presented with two choices that underestimate their time preference, say $\$ 40$ today or $\$ 39$ in 8 weeks. That is why we derived discount rates for each $m$ and $t$, using data from \citet{ifcher2011happiness}.

\subsection{Equalizing transaction costs}
At the very beginning of the questionnaire, participants were informed about the reward-claim process, in case they were chosen as the winner. The process was designed to equalize transaction costs and uncertainty associated with the payment. For example, if subjects believed that taking the reward at a later time would require some additional effort, then relative transaction costs of taking the money today would be lower, potentially affecting their responses to time preference questions. That is why they were instructed that the online payment would be made automatically, regardless of the date of the payment.

\subsection{Potential impact of COVID-19 on time preference}
The potential impact on time preference caused by COVID-19 crisis was also taken into account. In particular, we took into account two potential effects. We firstly considered this crisis as a health emergency, which causes people to stay at home across countries. If we compare the current health crisis to that caused by natural disasters, evidence from \citet{callen2015catastrophes} would suggest that people should become more patient in response to the crisis. Intuitively, when people are locked in their homes, they might be more willing to postpone their consumption in the present, because there is less choice of what to consume. However, if we consider the COVID-19 situation as an economic crisis, evidence from \citet{jetter2020becoming} suggests that worsening economic conditions (e.g. higher unemployment) makes people less patient. Because these two effects have opposing directions, we have assumed that they would on average cancel each other out.

\subsection{Trust Level Questionnaire}
After completing time preference questionnaire, subjects were asked five questions to test the success of trust-manipulation procedure. Each question was testing a different aspect of trust in anonymous individuals. Subject had to respond on a sliding scale, ranging from $-50$ to 50. These five questions are as follows:

\begingroup\itshape
\begin{enumerate}
    \item How much do you agree with the following statement: In general, you can trust people.

    \item How much do you agree with the following statement: Nowadays, you can't rely on anybody.

    \item  How much do you agree with the following statement: When dealing with strangers, it's better to be cautious before trusting them.

    \item  How much do you trust strangers you meet for the first time?

    \item Imagine you lost your wallet with your money, identification or address in your city/area and it was found by a stranger. How likely do you think your wallet would be returned to you? 
\end{enumerate}
\endgroup

Questions 1-4 were adopted from \citet{naef2009measuring}, while Question 5 was adopted from \citet{helliwell2010trust}. To avoid order effects, the sequence of these five questions was randomized. In Questions 1-3, five labels were provided as cues on the slider scale: \textit{disagree strongly (-50), disagree somewhat (-25), neutral (0), agree somewhat (25), agree strongly (50)}. In the analysis of results, the responses to 2 and 3 were multiplied by -1. In Question 4, four labels were provided as guidelines: \textit{I don't trust them at all (-50), I trust them very little (-25), I trust them quite a bit (25), I trust them a lot (50)}. In Question 5, four labels were again provided as cues: \textit{Not likely at all (-50), Not very likely (-25), Fairly likely (25), Very likely (50)}.

We should note that in surveys, such as General Social Survey (GSS) or World Values Survey (WVS), trust is measured by a person's binary agreement with the statement: \textit{Generally speaking, would you say that most people can be trusted or that you can't be too careful in dealing with people?} However, \citet{naef2009measuring} present evidence that trust in strangers can be more accurately measured by breaking down this statement into several questions.

\subsection{Certainty questionnaire}
Lastly, subjects were given four questions which were designed to measure their level of certainty about future outcomes. We conjectured that due to COVID-19 crisis, any question concerning people's certainty about the upcoming year might reflect numerous confounds, including political beliefs about how well the crisis is handled by national governments. Therefore, subjects were asked two sets of questions about more distant future, formulated in the same manner:

\begingroup\itshape
\begin{enumerate}
    \item Do you agree with the following statement: "In $t$ years, I will be better off than I am right now"

    \item How certain are you about your response?
    
\end{enumerate}
\endgroup

\subsection{Demographic questionnaire}
At the end of the survey, subjects were asked a series of demographic questions, which concerned their age, ethnicity, gender, education level, college major, and practice of religion. They were also provided the opportunity to leave an email address, in order to be considered for the lottery.

\subsection{Debriefing form}
After completing the experiment, subjects were provided with a debriefing form. The form detailed what was measured in each stage, and emphasized that the subjects were playing against a computer in the trust game part.

\section{Econometric Specifications}
\subsection{Measuring the Success of the Treatment procedure}

In analyzing the effect of playing the Trust Game on subjects' trust levels, we consider the fixed effects regression model of the form:

\begin{equation} \label{eq:1}
Trust_{i}= \beta H_{i}+\alpha_{i}+\sum_{K}^{N} \gamma_{K} I_{K}(k)+ \epsilon_{i}
\end{equation}

where $Trust_{i}$ is the measured level of trust in question $i \in\{1,5\}$. Regression analysis here is possible, because in each of the five questions, trust levels were measured as a continuous variable, where ${Trust_{i}} \in[-50,50]$. $H_{i}$ is the dummy of being in the High Trust treatment for question $i$. The question-specific intercept is denoted by $\alpha_{i}$, and the question-specific error term is denoted by $\varepsilon_{i}$. The model also includes demographics controls, where $I_{K}(k)$ is an indicator function, which takes the value of 1 if the subject belongs to the demographic category $K$. The number demographic categories is given by the interval $[\mathrm{K}, \mathrm{N}]$\footnote{In all regressions, demographic categories are: gender, age, ethnicity, education, college major, and religious practice.}.

We chose to conduct a fixed effects regression model for the following reasons. As we have explained in the experimental design section, each of the five questions in the trust level questionnaire measure a different aspect of trust in anonymous individuals. In addition, they are formulated in rather different ways - questions (1)-(3) ask subjects to state their level of agreement with a given statement, question (4) is a direct question, while question (5) requires subjects to estimate a probability. This led to us to expect that each of these questions might have different average responses, which we confirmed by testing the difference in mean responses. Therefore, we want to allow each question to have its own intercept, when estimating the average effect of the High Trust treatment on trust levels across these five questions. That is precisely the purpose of a fixed-effects regression model. We use OLS with robust standard errors to estimate the equation \ref{eq:1}

\subsection{Measuring the Effect of Trust on Time Preference}
To measure the effect of High Trust treatment on subjects' time preference, we used a regression model of the following form:

\begin{equation} \label{eq:2}
D=\beta H+\sum_{K}^{N} \gamma_{K} I_{K}(k)+ \epsilon
\end{equation}

where $H$ is the dummy of being in a High Trust treatment and $D$ is the discount rate. The model also includes demographics controls, where $I_{K}(k)$ is an indicator function, which takes the value of 1 if the subject belongs to the demographic category $K$. The number of demographic categories is given by the interval $[K, N]$. We use OLS with robust standard errors to estimate equation \ref{eq:2}. 

Our approach to estimating time preference is a standard exponential discounting model (Frederick et al., 2002), originally introduced by Samuelson (1937). It follows a literature of similar methods that more recently have been applied by \citet{benjamin2016religious}, \citet{REUBEN201563}, and \citet{burks2009cognitive}. Using our time preference questionnaire, we can determine the amount $x$, at which the subject is indifferent between receiving $x$ now, or receiving $p$ after $t$ weeks, where $p \in\{\$ 25, \$ 30, \$ 40\}$. Indifference implies that:

\begin{equation} \label{eq:3}
u(x)=D^{t} u(p)
\end{equation}

If we assume that utility is approximately linear, then taking the log of both sides yields:

\begin{equation} \label{eq:4}
\log x-\log p=t \log D
\end{equation}

Each subject was given 12 questions to measure their time preference. These questions formed 6 blocks, that enabled us to estimate 6 values of discount rates for the same individual. In each block, subjects were asked to choose between receiving an amount today and in the future. The combination of these amounts in the two questions is the same. The difference between the two questions in a block is the time distance $t$. For every block, $x$ is the amount at which the individual switched to the present-day payment and $t$ is time delay in weeks.

\subsection{Measuring the Effect of Trust on Certainty about Future Outcomes}
To estimate the effect of High Trust treatment on subjects' certainty about future outcomes, we used a regression of the following form:

\begin{equation} \label{eq:5}
Certainty =\beta H+\sum_{K}^{N} \gamma_{K} I_{K}(k)+\varepsilon
\end{equation}

where $H$ is the dummy of being in a High Trust treatment and Certainty is a measure of subjects' certainty about future outcomes, Certainty $\in[0,100]$. The model also includes the same demographics controls, which are used in other regression models. We use OLS with robust standard errors to estimate equation \ref{eq:5}. 

\section{Results}
\subsection{Demographic Characteristics of Subjects}

\begin{table}[ht]
    \centering
    \caption{Demographic statistics}
    \begin{tabular}{c}
        \includegraphics[scale = 0.9, clip, trim=2.5cm 17cm 2cm 2cm]{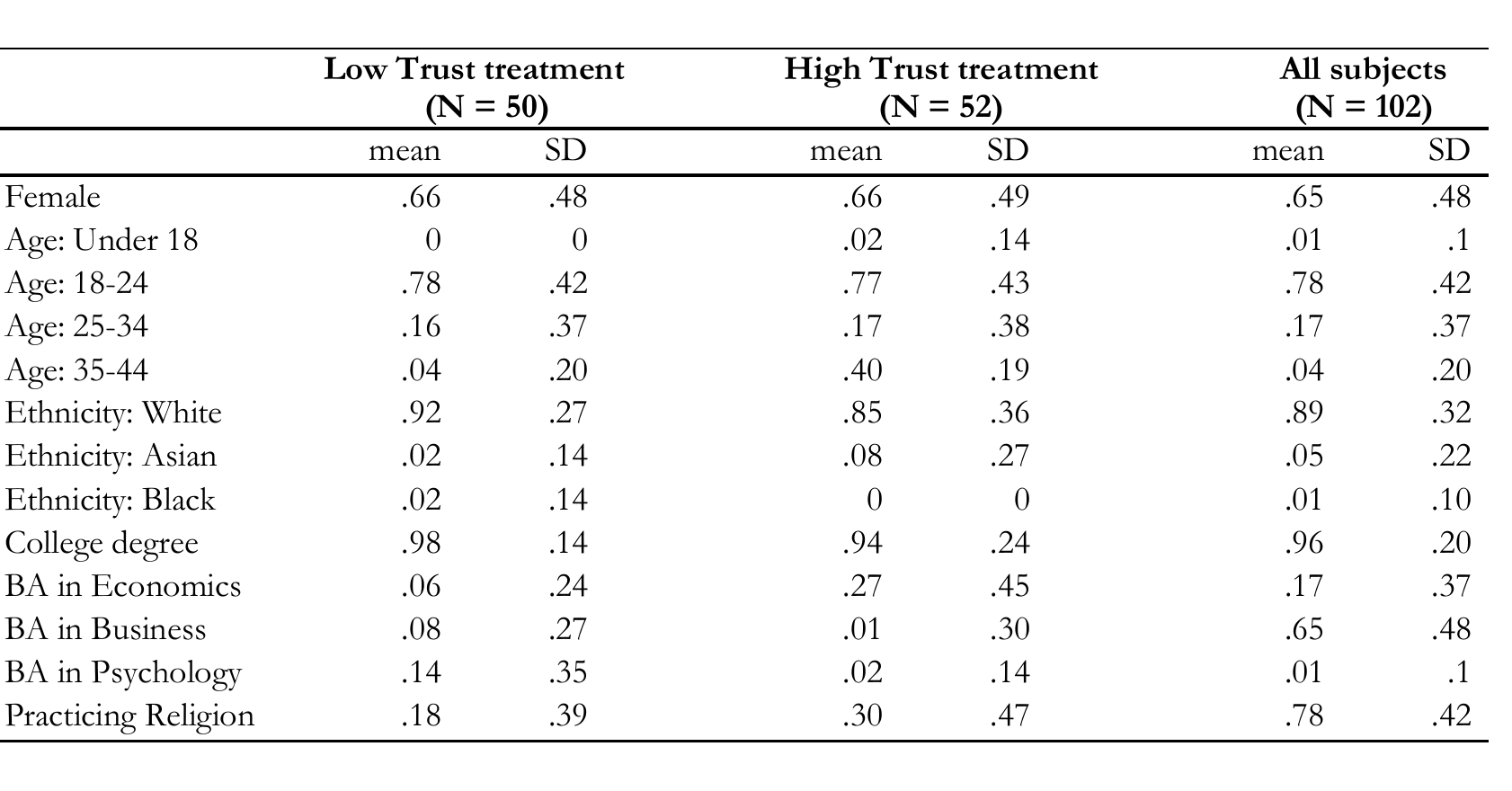}
    \end{tabular}
    \label{tab:}
\end{table}

Table 2 presents demographic characteristics of subjects in the experiment. We find significant difference between the values of two demographic characteristics of the two treatment groups. Therefore, even though the assignment to a treatment group was completely random during the experiment, we cannot conclude it is completely random ex post. First, the number of students who are pursuing or have pursued a Bachelor's degree in Economics is significantly different (two-sided t-test yields $\mathrm{p}=0.0043$ ): 3 subjects in Low Trust treatment, and 14 subjects in High Trust treatment. Moreover, the number of students who are pursuing or have pursued a Bachelor's degree in Psychology is significantly different (two-sided t-test yields $\mathrm{p}=0.0233$ ): 7 subjects in High Trust treatment, and 1 subject in Low Trust treatment. These two groups are important primarily because they might have had prior exposure to the Trust game, which might have had an effect for the success of the treatment. However, we have controlled for these variables in our regression models, so they should not affect the validity of our results. As a precaution, in the analysis that follows, we have also included a regression with two interaction variables: BA in Economics $\mathrm{x}$ High Trust treatment, and BA in Psychology $\mathrm{x}$ High Trust treatment.

\subsection{High Trust Treatment Increases Trust Levels}

We find evidence that High Trust treatment significantly increases trust levels, which are presented in Table 3 below. This effect is stable regardless of specification. In column 1 , we find that the effect of High Trust treatment on trust levels holds without controlling for demographic variables. On average, playing against the High Trust algorithm instead of the Low Trust algorithm during the trust game increases trust levels by around $4.5$ points, where they were measured on a scale from $-50$ to 50 . The result is significant at $5 \%$ level. In columns 2-4, controls are added for gender, age, ethnicity, education, religious practice, and college major. When demographic controls are added, treatment becomes significant at $1 \%$ level, increasing trust levels by around 6 points.

\begin{table}[ht!]
    \centering
    \caption{Estimates from equation \ref{eq:1}}
    \begin{tabular}{c}
        \includegraphics[scale = 0.9, clip, trim=2.5cm 5.5cm 2cm 2cm]{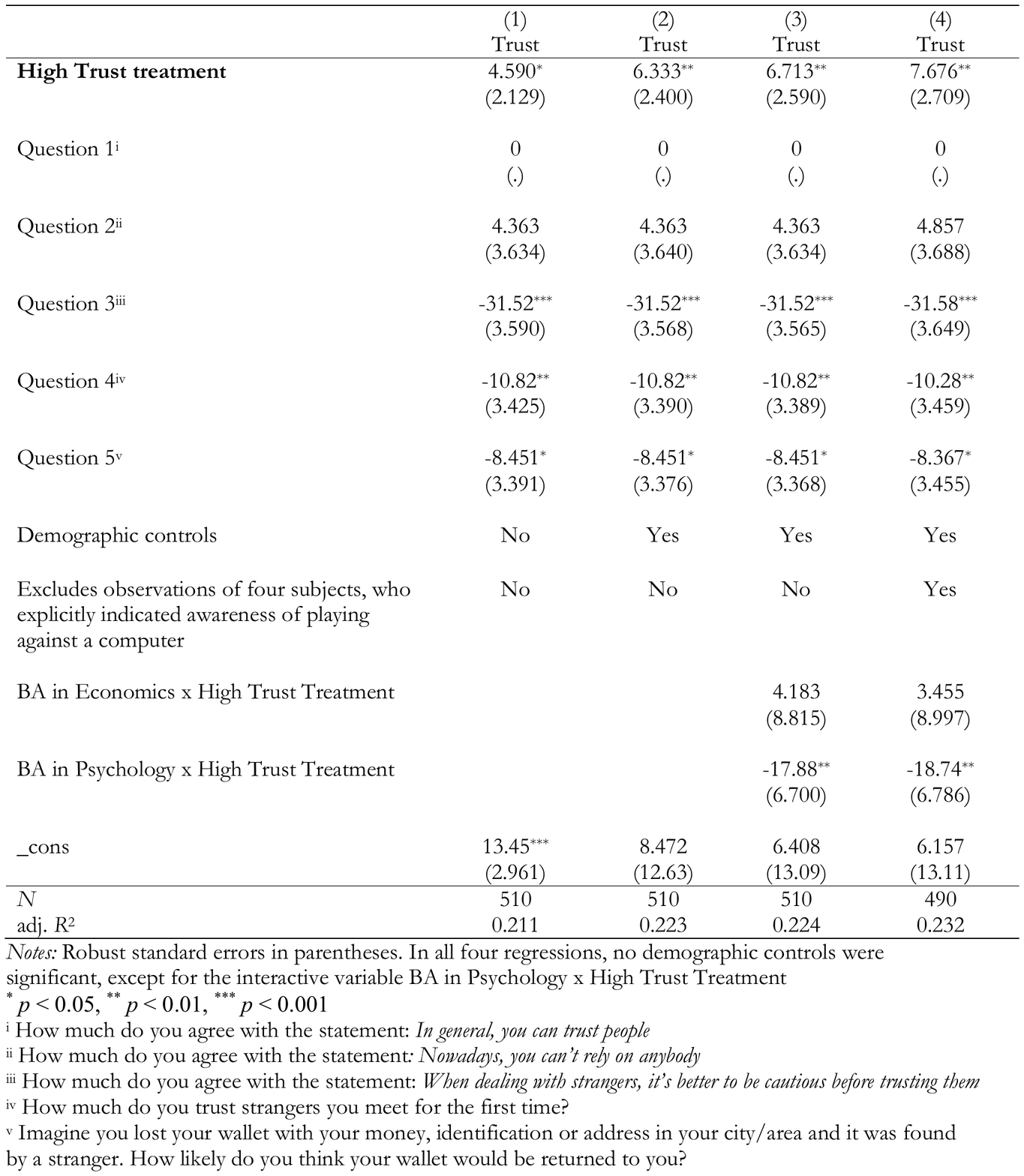}
    \end{tabular}
    \label{tab:}
\end{table}

In column 3 , the analysis of column 2 is repeated, adding interactive variables of BA in Economics x High Trust treatment, and BA in Psychology x High Trust treatment. These variables are included, because there is a statistically different number of participants who are pursuing or have pursued a Bachelor's degree in Economics or Psychology between the two treatment groups. We can see a slight increase in both the coefficient and the standard error of the High Trust treatment dummy. This effect might be due to the result that the interactive term BA in Psychology $\mathrm{x}$ High Trust treatment is significant at $1 \%$ level. 

Finally, in column 4, the regression of column 3 is repeated, but observations of four subjects from the experiment are excluded. These four subjects had explicitly indicated their awareness of playing against a computer in the treatment procedure in the feedback form at the end of the experiment. As we have discussed earlier, evidence suggests that people evoke their trust beliefs in the Trust game more strongly, if they believe they are playing against a real individual. With the restricted sample, the effect of the treatment increases further. We can also see that the effect of the treatment is significant on every individual question in the trust questionnaire, except for Question 3.

We also find a significant difference between mean responses to each of the five questions, taking both treatment groups together. In Questions (1)-(5), the mean trust level is respectively $15.8,20.2,-15.7,5.0,7.3$.

In summary, the first finding of this experiment is that High Trust treatment significantly increases trust levels, compared to the Low Trust treatment. This effect is robust to a number of econometric-specification checks.

\subsection{There is No Effect of High Trust treatment on Time Preference}
We find no significant evidence that High Trust treatment affects time preference. Out of the six regressions presented in Table 4, we find a significantly negative effect of High Trust treatment on time preference in one of them, which does control for demographic variables. Once demographic variables are included, the significance disappears. Even if statistically insignificant, the results are somewhat surprising - in all six regressions, the coefficient of High Trust treatment is negative.

\begin{table}[ht!]
    \centering
    \caption{Estimates from equation \ref{eq:2}}
    \begin{tabular}{c}
        \includegraphics[scale = 0.9, clip, trim=2cm 14.5cm 2cm 3.5cm]{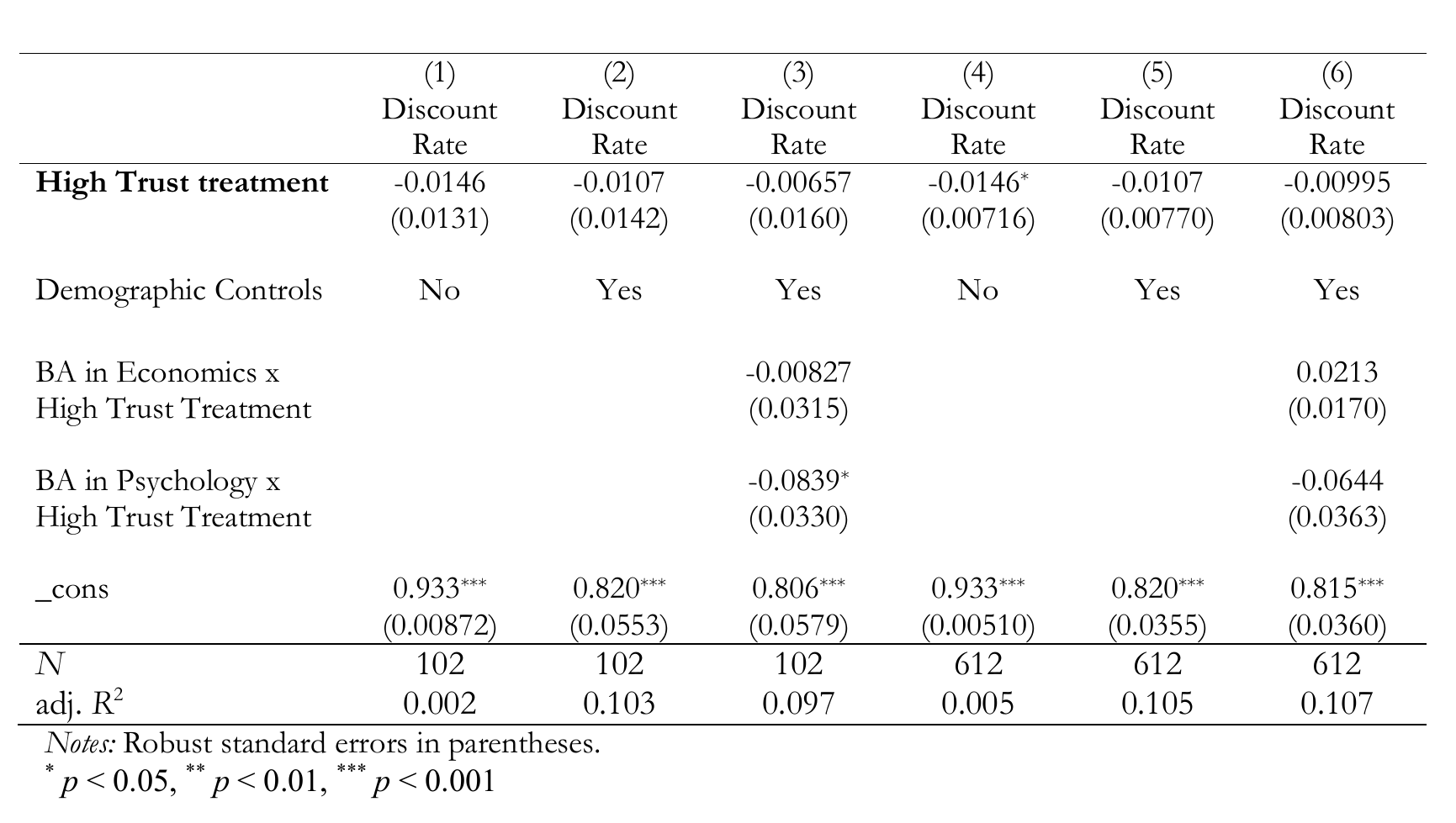}
    \end{tabular}
    \label{tab:}
\end{table}

In columns $1-3$, time preference is measured as the average of the six discount rates we have calculated for every individual. Column 1 presents regression results that do not control for demographic variables. The effect of High Trust treatment is negative, but insignificant at $5 \%$ level. In column 2 , the analysis of column 1 is repeated, adding demographic controls. The effect remains insignificant and negative. The model in column 3 adds two interactive variables, Economics $\mathrm{x}$ High Trust treatment, and BA in Psychology x High Trust treatment. The effect of High Trust treatment on time preference remains insignificant. However, the interactive term BA in Psychology $\mathrm{x}$ High Trust treatment is significant at $5 \%$ level.

In columns 4-6, time preference for each individual is measured as six different discount rates, which we have calculated for each combination of the monetary amount and time horizon. For this reason, the number of observations increases from $\mathrm{N}=102$ to $\mathrm{N}=612$. Column 4 presents regression results that do not control for demographic variables. The effect of High Trust treatment is negative and significant at $5 \%$ level. There are two reasons why we do not draw any conclusions from this result. First, once demographic controls are included in columns 5-6, the significance disappears. Evidence suggests that time preference tends to vary with demographic characteristics \citep{harrison2002time}, and we have already shown that our random assignment process did not produce completely random samples from a demographic point of view. For this reason, including demographic variables should yield more accurate results. Moreover, it is important to look at the adjusted R-squared value in column 4, which is $0.005$. In both absolute terms, and compared to regressions that include demographic variables in columns 5 and 6 , this value is considerably small: High Trust treatment explains around $0.5 \%$ of variation in time preference. Once demographic variables are included, treatment explains over $10 \%$ of the same variation.

In column 5 , the analysis of column 4 is repeated, adding demographic controls. The effect remains insignificant and negative. Column 6 adds two interactive variables, Economics $\mathrm{x}$ High Trust treatment, and BA in Psychology x High Trust treatment. The effect of High Trust treatment on time preference remains insignificant.

\subsection{There is No Effect of High Trust Treatment on Certainty about the Future}

We find no significant evidence that High Trust treatment affects certainty about future outcomes, presented in Table 5. Even though the results are insignificant, it is somewhat surprising that both coefficients are negative. In both columns, levels of certainty are measured as individual responses to each of the two certainty questions. For every subject, we therefore have 2 observations, leading to $N=204$. In column 1 , regression does not include demographic variables. Adding demographic controls in column 2 does not make the results significant, but it raises the value of adjusted R-squared from $0.000$ to $0.125$. As we will discuss in the next section, even if this result was significant, it would be difficult to make strong conclusions due to the potential impact of COVID-19 on the levels of certainty.

\begin{table}[ht!]
    \centering
    \caption{Estimates from equation \ref{eq:5}}
    \begin{tabular}{c}
        \includegraphics[scale = 0.9, clip, trim=2cm 17.5cm 2cm 2.25cm]{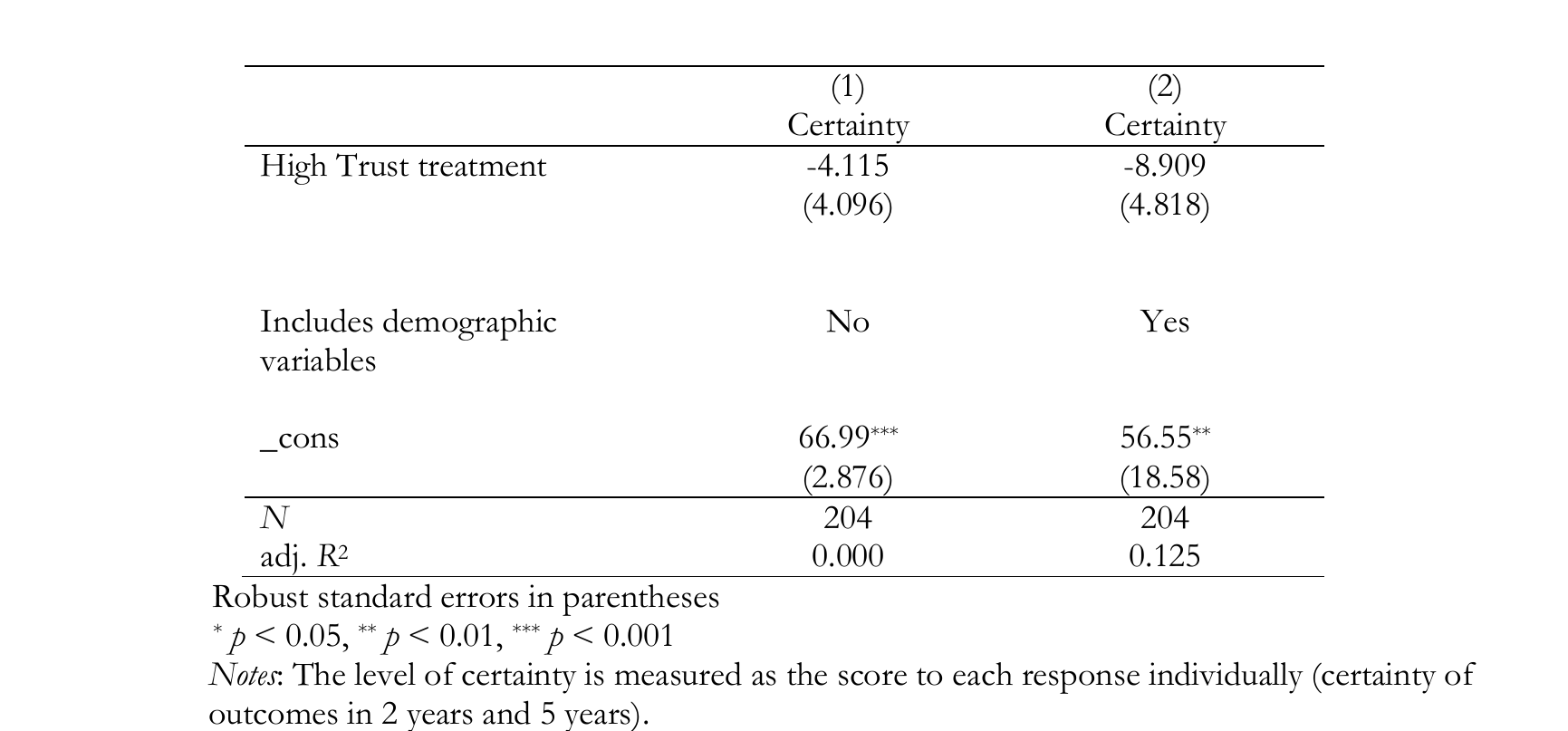}
    \end{tabular}
    \label{tab:}
\end{table}

\section{Discussion}
Our research presents two main findings. First of all, trust has no significant effect on time preference or levels of certainty. Second, it is possible to manipulate people's short-term levels of trust for experimental purposes. The first finding sheds light on the relationship between trust and patience. Given that both variables seem to have an interconnected relationship with economic growth, it helps us understand how trust and patience affect each other, which has not been studied much in the literature. More specifically, if trust has little causal effect on patience, it is likely that patience might have an effect on trust, or that they are both related to a third confound. The second finding opens the door for further investigations into causal effects of trust on economic outcomes - a topic that is quite integral to the understanding of the determinants of economic growth, yet has been studied in very few experiments. We would like turn to several important shortcomings of this experiment.

\subsection{Internal Validity of the Experiment}
\subsubsection{Measuring time preference}
First of all, there are potential shortcomings in analysing the level of time preference in this experiment. The main problem here is a possible confounding variable - the effect of COVID-19 health crisis on subjects' time preference. As we have discussed in the section on experimental design, time preference questions were constructed in a way that would maximize the difference of answers from subjects with distinct levels of discounting. However, the choice design used discount rates that people exhibit in normal conditions, rather than a situation of a severe health and economic crisis. We have conjectured two potential effects of COVID-19 emergency on people's discounting rates - individuals might become more patient, because there are fewer opportunities to spend the reward today compared to the future, or they might become less patient, because worsening economic conditions usually push people to consume more today and save less for tomorrow. Because these mechanisms have contrasting effects on time preference, we have assumed that on average, they will cancel out. However, one of them might have been stronger than the other, or unequally distributed across the two treatment groups.

Moreover, because participants were recruited using personal social media account, our sample likely suffers from selection bias. Most of the subjects were either direct friends of the author, friends of friends, or other students at Columbia University. This is problematic not only because it's an unrepresentative sample. Feedback after the experiment suggests that a few friends might have chosen to take a lower monetary reward today over a higher one in the future, not out of a personal preference, but rather because they thought the experiment was funded by the author himself, and they wanted to minimize his expenses. While that is a very considerate gesture from one perspective, it might have had a problematic effect on time preference data. If we assume that subjects from High Trust treatment were more likely to behave in this way (increased trust could trigger other prosocial preferences), this might serve as one of the many explanations for why the coefficient of High Trust treatment on time preference was actually negative, even though insignificant.

Lastly, there is a one validity issue that concerns the time preference survey itself. The results of the experiment indicate that the mean discount rate for Low Trust treatment was slightly higher than that of the High Trust treatment. However, the opposite is true for the variance of responses - subjects in the High Trust treatment had a higher variance of their binary responses. It is true that the reason why discount rates are slightly higher in the Low Trust group is that its subjects chose one type of a payment more often. At the same time, this might reflect another trend: Low Trust subjects might have been more consistent in their choices, because they were actually less patient, and wanted to get through the 12-time preference questions as quickly as possible. For the purpose of speed, a rule of thumb of choosing the same response in each question might have been quite useful, yet it might also be falsely interpreted as a feature of consistency, associated with a higher level of patience. Feedback from one participant suggests this might have been the case for some subjects.

\subsubsection{Measuring certainty}
Even if we had obtained any significant results about the effects of treatment on people's certainty about future outcomes, it would be very difficult to draw strong conclusions from them. The main reason for that is again the COVID-19 situation, which might have had a profound impact on the levels of certainty about the future. Moreover, certainty levels at the moment might reflect political biases as much as personal beliefs, since at least in the United States, some divergence in the way people perceive this emergency reflects the level of political support for the work of the current administration of the US government\footnote{See, for example: D. Roberts, "Partisanship is the strongest predictor of coronavirus response", published in Vox, 31 March, 2020.}. 

Moreover, the experiment could be strengthened by a more rigorous procedure to test the certainty levels themselves. It would be interesting to test two interrelated variables: the accuracy of the predictions of an outcome, as well as the associated sense of confidence of this prediction. \cite{meyniel2015confidence} have introduced an experimental design for testing precisely these two variables. In their experiment, subjects estimate transition probabilities between two visual or auditory stimuli in a changing environment, and report their mean estimate and confidence in this report\footnote{For the suggestion of this study, I am indebted to Arthur Prat-Carrabin from Columbia University’s Cognition and Decision Lab.}. We therefore invite researchers to explore the relationship between trust and certainty further.

\subsubsection{Treatment: trust manipulation procedure}
There are also several potential issues that concern the internal validity of the treatment procedure. The most important problem is that some subjects were aware they were playing with a computer, instead of a real person. As we have discussed before, evidence suggests that trust beliefs are only active during the Trust game, when when subjects think they are playing against another person, rather than a computer. Therefore, the effect of High Trust treatment on trust levels is potentially stronger. What is important to emphasize again is that in a lab setting, there would be no inherent need to use deception - it was used only because the experiment had to be conducted online.

\subsection{Implications: Moving Forward}
We would like to end by emphasizing three main takeaways. First of all, most of internal validity problems discussed above could be solved without increasing the budget of the experiment, or its sample size. Instead, it would suffice to conduct the experiment in a lab with anonymous subjects, at a time when the current public health crisis will be largely resolved. For this reason, we believe it would be worthwhile to replicate the experiment again in the future, when such conditions can be met. Second, the success of the trust-manipulation treatment procedure in this experiment opens ample opportunities for further research in the field of economics of trust. Understanding the causal relationship between trust and other economic outcomes might shed light on the precise mechanisms by which trust affects economic growth, which are not yet clear at the moment. More broadly, the treatment could be useful for any experiment that examines the causal effect of trust even outside of the field of economics, in areas like psychology or sociology. As we have discussed before, there is no need to use deception to change people's short-term beliefs of trust, when using the treatment outlined in this experiment if it is conducted in a lab setting. 

Above all, we hope that this paper will serve as an invitation for others to investigate the numerous interesting links between trust, certainty, time preference, and economic growth.

\section{Appendix}
\begin{table}[ht!]
    \centering
    \caption{Strategies of the Algorithm as Participant A}
    \begin{tabular}{c}
        \includegraphics[scale = 0.9, clip, trim=2cm 20cm 2cm 2.25cm]{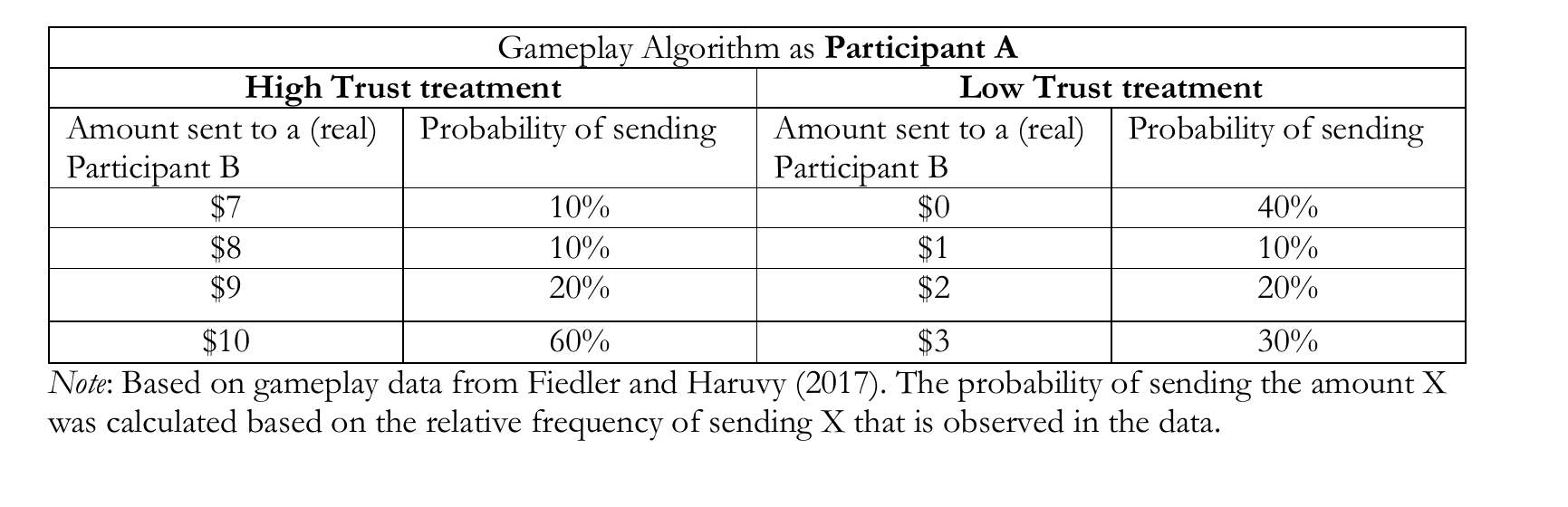}
    \end{tabular}
    \label{tab:}
\end{table}

\begin{table}[ht!]
    \centering
    \caption{Strategies of the Algorithm as Participant B}
    \begin{tabular}{c}
        \includegraphics[scale = 0.9, clip, trim=2cm 14.5cm 2cm 2.25cm]{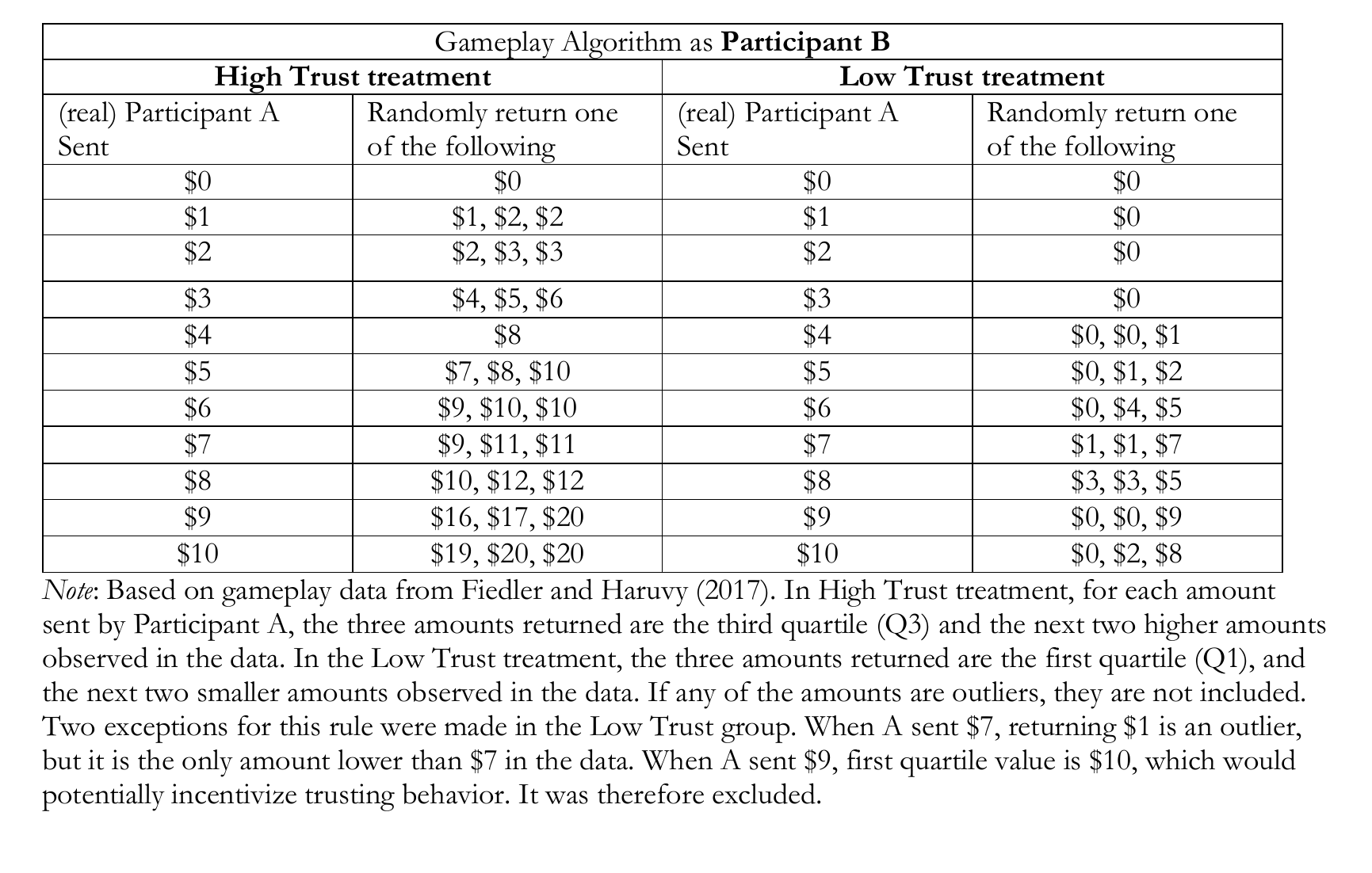}
    \end{tabular}
    \label{tab:}
\end{table}

\pagebreak
\pagebreak

\end{document}